\documentstyle[12pt,twoside]{article}
\input{epsf}
\begin{document}
\setlength{\baselineskip}{12pt}
\hoffset 0.65cm
\voffset 0.3cm
\bibliographystyle{normal}

\vspace{48pt}
\noindent
{\bf Temperature Dependence of Anchoring Energy of MBBA on 
$SiO$-Evaporated Substrate}

\vspace{48pt}
\noindent
M.V. KHAZIMULLIN, A.P. KREKHOV and Y.A. LEBEDEV

\noindent
Institute of Molecule and Crystal Physics, Russian Academy of Sciences, 450025 Ufa, Russia

\vspace{36pt}
\noindent
The temperature behaviour of the anchoring energy of nematic 
liquid crystal MBBA at the substrates with oblique $SiO$ evaporation 
and director orientation at the interface have been 
studied by means of magnetic Fr\'eedericksz transition technique.
The temperature dependence of the coefficients in the 
phenomenological model of surface anchoring energy has been determined.

\vspace{24pt}

\vspace{36pt}
\baselineskip=1.2 \baselineskip

\noindent
{\bf INTRODUCTION}
\vspace{12pt}

\noindent
The equilibrium director orientation in a nematic liquid crystal (NLC) layer strongly 
depends on the anchoring properties of the bounding substrates.
In the absence of an external field the confining surfaces impose the 
orientation in the bulk of nematic layer.
If an external field is present the director distribution results from the competition
between the surface and bulk torques.
Study of the orientational behaviour in an external field provides the information about 
anchoring properties of NLC at the bounding substrates \cite{Rapn,Cognard}.
Besides, a useful tool for the investigation of interfacial phenomena 
is the temperature-induced
orientational transitions which have been found at some interfaces 
\cite{Lang,Lisi,Lgrw}.

Recently the temperature-induced orientational transitions planar $\to$ tilted $\to$ homeotropic close
to the clearing point (nematic - isotropic transition) have been found at the interface 
between MBBA and a $SiO$-evaporated substrate \cite{Cryst}. 
To describe this orientational behaviour the phenomenological model of surface energy
similar to \cite{Sull,Barb} taking into account the substrate micro relief has been suggested. 

In this paper we present the results of experimental investigations of the orientational
behaviour of the nematic liquid crystal MBBA in a magnetic field on the same interface as in \cite{Cryst} 
in the temperature range where the orientational transitions occurred. 
The director tilt angle at the substrate as a function of temperature has been measured 
and compared with theoretical predictions. 
The parameters of the phenomenological model are determinated and the
temperature dependence of surface energy is found.

\vspace{24pt}
\noindent
{\bf BASIC EQUATIONS}
\vspace{12pt}

\noindent
We consider a nematic liquid crystal layer of thickness $l$ confined between two
identical plates at $z=0$ and $z=l$.
Weak anchoring is provided by the substrates with micro relief (grooved
surface).
The substrate with micro relief is characterised by the grooves axis
(along $x$ axis) and normal vector.
Then the phenomenological expression for the surface anchoring energy  $f_s$ for such substrate 
can be obtained using the expansion
in terms of the nematic tensor order parameter $Q_{ij}=S(T)[n_i n_j -1/3 \delta_{ij}]$ up to
second order in the scalar order parameter $S(T)$ (see \cite{Cryst} for details)
\begin{equation}
\label{senr}
f_{s} = C(S)\cos^2\theta_0 + D(S)\cos^4\theta_0 \;,
\end{equation}
with 
\begin{equation}
\label{CD}
C(S) = a S + b S^2 \;,\;\; D(S) = d S^2 \;,
\end{equation}
where $\theta_0$ is the angle between the director at the substrate and
the $x$ axis and 
$a$, $b$, $d$ are the temperature independent model parameters which depend on 
the micro relief properties and the nematic - substrate interaction.
If there is no external field the director distribution in the nematic layer is defined by 
the surface angle $\theta _0$, which can be found by minimisation of the surface energy 
$f_s$ with respect to $\theta _0$. 
Depending on the model parameters ($a$, $b$ and $d$) one could have different
scenarios for the temperature behaviour of the surface angle $\theta_0$.
For example, $a=0$, $b<0$, $d>0$, $b+2d>0$ describes a temperature independent pretilt, or, 
if $a>0$, $b<0$, $d>0$, $a+b<0$, $a+b+2d>0$,
one has a tilted $\to$ homeotropic transition similar to that found in \cite{Lang}.
Here we focus on the case of $a>0$, $b<0$, $d>0$, $a+b<0$, $a+b+2d<0$ when the temperature-induced
orientational transition planar $\to$ tilted $\to$ homeotropic takes place.
Minimisation of surface energy (\ref{senr}) gives the temperature behaviour of surface angle 
\begin{equation}
\label{th0S}
\cos^2\theta_0 = \frac{S_p}{S_p-S_h}(1-\frac{S_h}{S}) \;,
\end{equation}
with transition points planar $\to$ tilted ($S_p$) and tilted $\to$ homeotropic ($S_h$)
\begin{equation}
\label{SpSh}
S_p = -\frac{a}{b+2d} \;,\;\; S_h = -\frac{a}{b} \;.
\end{equation}
The temperature dependence of the surface angle $\theta_0$ can be found 
from Eq.(\ref{th0S}) using the 
experimental data for the scalar order parameter $S(T)$ \cite{parm}.

In the case when a magnetic field ${\bf H}$ is applied along the $z$ axis,
the director distribution can be obtained by minimisation of the
total free energy (per unit area)
\begin{equation}
\label{gener}
F = \frac12 \int_{0}^{l} \left[ K(\theta ) \left( \frac{d\theta}{dz} \right)^2 
- \chi_a H^2 \sin^2\theta \right] dz + f_{s 1} + f_{s 2} \;,
\end{equation}
where $K(\theta )=K_1\cos^2\theta + K_3\sin^2\theta$,
$K_1$ and $K_3$ are elastic constants for ``splay'' and ``bend'' deformations respectively,
$\chi _a$ is the diamagnetic anisotropy and $f_{s i}$ are the surface 
energies for the lower ($i=1$) and upper ($i=2$) substrates.
After the standard procedure one gets the equation for the 
director profile (first integral of Euler-Lagrange equation)
\begin{equation}
\label{deriv}
\frac{d\theta}{dz} = H \sqrt{\chi _a\frac{\cos^2\theta - \cos^2\theta_m}{K(\theta)}}
\end{equation}
and boundary condition (in the case of identical substrates
$f_{s 1} = f_{s 2} = f_s$)
\begin{equation}
\label{torq}
\frac{d f_s}{d\theta _0} = H \sqrt{\chi _aK(\theta _0)(\cos^2\theta _0 - \cos^2\theta _m)} \;,
\end{equation}
where $\theta _m$ is the angle in the midplane of the nematic layer and the symmetry of the
solution with respect to the midplane has been used ($d\theta/dz=0$ at $z=l/2$).
Integrating (\ref{deriv}) from $\theta _0(z=0)$ to $\theta _m(z=l/2)$
gives the relation between $\theta _m$ and $\theta _0$
by 
\begin{equation}
\label{thM}
\frac{\pi }{2}\frac{H}{H_{F}}=\int _{\theta _{0}}^{\theta _{m}}P(\theta )d\theta \; , \;\;
P(\theta)=\sqrt{\frac{1+\eta \cos^2\theta}{\cos^2\theta - \cos^2\theta_m}} \; ,
\end{equation}
where $H_F=(\pi/l)\sqrt{K_1/\chi_a}$ is the Fr\'eedericksz transition field for strong planar 
anchoring and $\eta =K_3/K_1-1$.

Using expressions (\ref{senr}), (\ref{CD}), (\ref{SpSh}) and (\ref{torq}) the temperature 
behaviour of the surface angle $\theta_0$ in the presence of a magnetic field can be derived in the following form
\begin{eqnarray}
\label{torqb}
-\sin(2\theta_0) b S \{ (S-S_h) - (1 - \frac{S_h}{S_p}) S \cos^2\theta _0\}
\nonumber \\
  = H \sqrt{\chi _a K(\theta _0)(\cos^2\theta _0 - \cos^2\theta _m)} \;.
\end{eqnarray}
The values of $S_p$, $S_h$ can be easily found from the experimental 
data on the planar $\to$ tilted $\to$ homeotropic 
transition in zero magnetic field.
Measuring the temperature dependence of the surface angle $\theta_0$ at different values of
the magnetic field one can verify the phenomenological model (\ref{senr}), (\ref{CD}) and determine the 
model coefficients (which should be independent of the magnetic field).

\vspace{24pt}
\noindent
{\bf EXPERIMENTAL}
\vspace{12pt}

\noindent
The experimental cell consists of two glass plates with mylar spacers 
of thickness $\sim 35$ $\mu$m
filled by nematic liquid crystal MBBA.
The inner surface of the confining plates was covered by thin layer of 
$SiO$, which was vacuum evaporated under the
angle $60^{\circ}$ with respect to the surface normal.
At room temperature one obtains homogeneous planar orientation (along
$x$ axis).
The cell was mounted in the hot stage and demonstrated the temperature-induced orientational 
transition planar $\to$ tilted $\to$ homeotropic under heating.
The accuracy of the temperature measurements was better than $10$ mK.
To study the orientational behaviour of NLC in a magnetic field the experimental
setup consisting of an electromagnet with maximum field $12$ kOe and $12$
mm gap between the poles has been designed.
The stability of the magnetic field was $\pm 10$ Oe.
The intensity of transmitted light as a function of temperature $I(T)$
at the fixed values of magnetic field has been measured by a photometer MPM-100 (Zeiss)
(light source: He-Ne laser, $\lambda=632$ nm).
All measured signals (magnetic field, temperature, and transmitted intensity) were processed 
by a computer ADC-card.

For the case of cross polars and $x$ axis at $45^{\circ}$ to the polarisers one has for the
transmitted light intensity
\begin{equation} \label{ints}
I = I_0 \sin^2\frac{\delta}{2} \;,\;\; \delta =\frac{2\pi}{\lambda}\int _{0}^{l} R(\theta)dz
\end{equation}
with
\begin{equation} \label{R}
R(\theta) = \frac{n_o n_e}
      {\sqrt{n_o^2\cos^2\theta + n_e^2\sin^2\theta}} - n_o \;.
\end{equation}
Here $\delta$ is the phase difference; 
$\lambda$ is wavelength of light; $n_o$, $n_e$ are the ordinary and 
extraordinary refractive indices.
Since $\delta=\delta(T)$ is monotonically decreasing to zero during 
planar $\to$ tilted $\to$ homeotro\-pic 
transition, it can be easily obtained from the experimental data on the temperature dependence
of the transmitted light intensity.
Then, taking into account (\ref{deriv}), (\ref{ints}) and (\ref{R}) one obtains the relation between
the phase difference and surface angle
\begin{equation} \label{delt-th0}
\delta =\frac{4 l}{\lambda}\frac{H_F}{H}
         \int _{\theta _m}^{\theta _0}R(\theta)P(\theta)d\theta \;.
\end{equation}
Using the temperature dependence of the material parameters $K_1$, $K_3$ and $\chi_a$ of MBBA 
\cite{parm}, expressions (\ref{delt-th0}), (\ref{thM}) allow to calculate the temperature
dependence of the surface angle $\theta_0(T)$ from the experimental data on the phase difference
$\delta(T)$ for the different values of magnetic field.

\vspace{24pt}
\noindent
{\bf RESULTS}
\vspace{12pt}

\noindent
In Fig.~1 the typical dependence of the transmitted light intensity on the reduced temperature 
$\tau =T/T_c$ for zero magnetic field close to the clearing point is shown.
The temperature dependence of the surface angle $\theta_0$ derived from these experimental data
is plotted in Fig.~2 (triangles).
The corresponding transition points planar $\to$ tilted ($\tau_p$) and 
tilted $\to$ homeotropic ($\tau_h$) are shown
at Fig.~1 by arrows.
The dependence $\theta_0(\tau)$ for zero magnetic field has been fitted
by Eq.(\ref{th0S}) and 
the transition points have been determined: $\tau_p=0.9961$ and $\tau_h=0.9984$. 

\begin{figure}
\begin{center}
\vspace*{-0.8cm}
\hspace*{-1cm}
\epsfxsize=10cm
\epsfbox{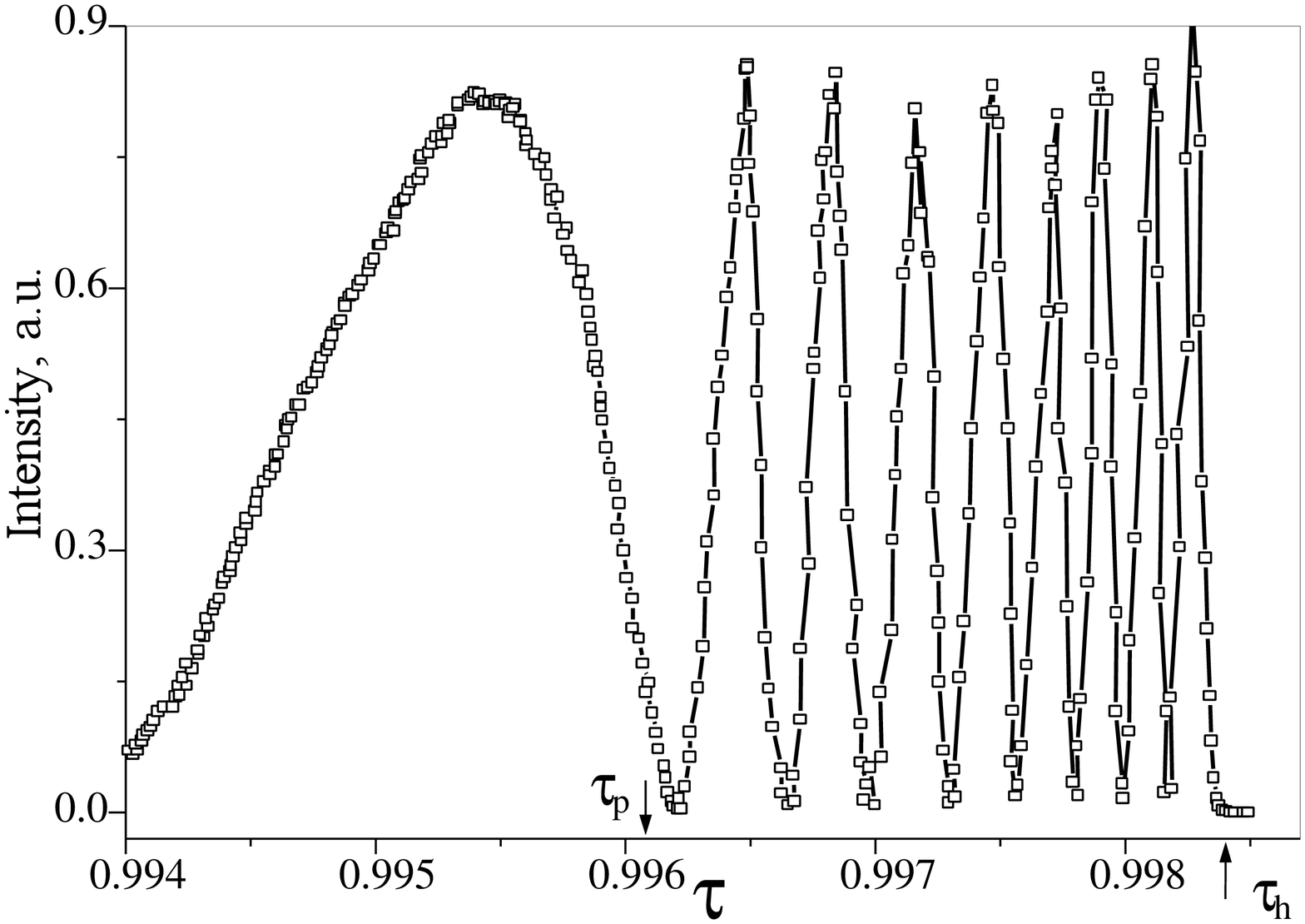}
\end{center}
\vspace*{-0.5cm}
\caption{Typical temperature dependence of the transmitted light intensity in the range of 
the orientational transition (thickness $l=36.9$ $\mu$m)}
\label{fig1}
\end{figure}
\begin{figure}
\begin{center}
\vspace*{-0.5cm}
\hspace*{-1cm}
\epsfxsize=10cm
\epsfbox{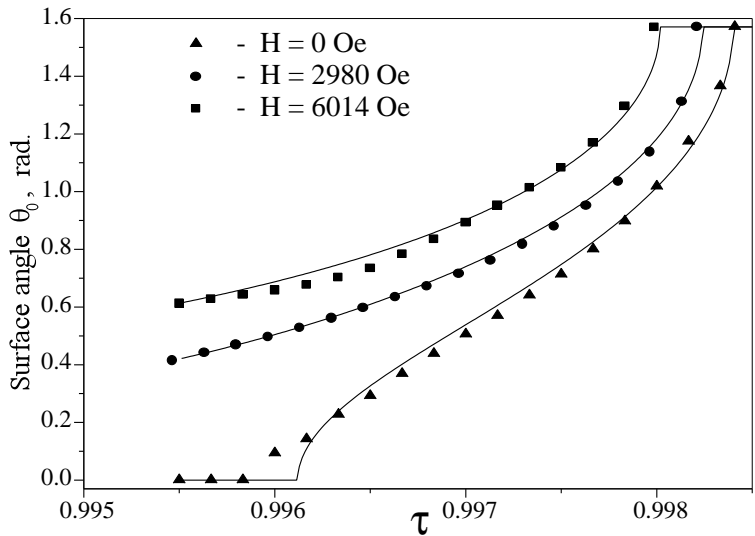}
\end{center}
\vspace*{-0.5cm}
\caption{Experimental (symbols) and theoretical (solid lines) 
temperature dependence of the
surface angle $\theta_0$ in a MBBA layer (thickness $l=36.9$ $\mu$m).}
\label{fig2}
\end{figure}

In the presence of a magnetic field the surface angle deviates from the initial value $\theta_0=0$
and then monotonically increases up to $\theta_0=\pi/2$ with increasing temperature (Fig.~2).
Note, that the temperature of the tilted $\to$ homeotropic transition is decreased with increasing 
magnetic field.
The solid lines in Fig.~2 show the behaviour of the surface angle calculated from the theoretical 
model for corresponding magnetic fields.
Using in Eq.(\ref{torqb}) the values of $\tau_p$ and $\tau_h$ found for
zero magnetic field,
one obtains $b=-0.055$ erg/cm$^2$ by fitting to the experimental data  
for $\theta_0(\tau)$ for different values of magnetic field.
In a wide range of magnetic field in the experiments the coefficient $b$ was found to be
independent of $H$.

\begin{figure}
\begin{center}
\vspace*{-0.8cm}
\hspace*{-1cm}
\epsfxsize=10cm
\epsfbox{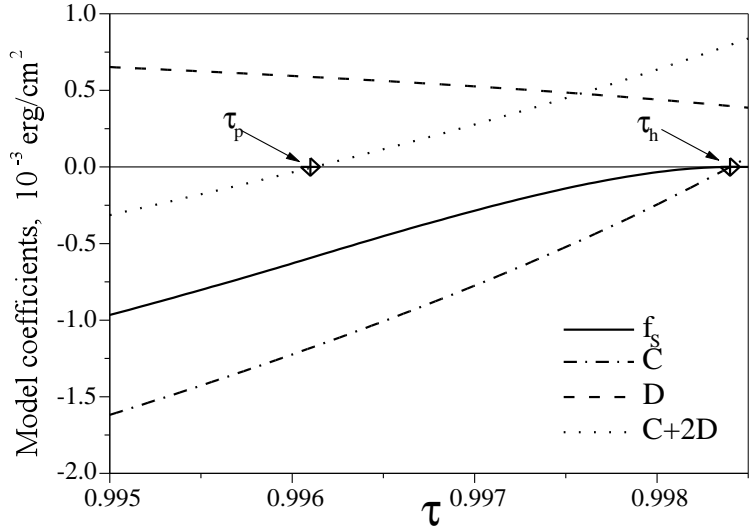}
\end{center}
\vspace*{-0.5cm}
\caption{Temperature dependencies of the surface energy $f_s$ and coefficients $C$, $D$, $C+2D$
in the model (\protect\ref{senr}).}
\label{fig3}
\end{figure}

Finally, in Fig.~3 the temperature dependence of the coefficients 
$C(\tau)$, $D(\tau)$ and $C(\tau)+2D(\tau)$ are presented together
with the surface energy $f_s$ [Eq.(\ref{senr})].
One sees that $C(\tau)$ changes sign at the tilted $\to$ homeotropic 
transition point ($\tau_h$) and $C(\tau)+2D(\tau)$ at the 
planar $\to$ tilted transition point.
The surface energy $f_s$ is quite small and decreases to zero at the clearing point.
At temperature $T=26$ $^{\circ}$C one gets for the surface energy 
$8 \cdot 10^{-3}$ erg/cm$^2$.

It should be noted, that MBBA is a rather unstable substance, which 
can have an influence on the determination of model coefficients.
For estimation of possible errors the coefficients $C$, $D$ were 
calculated taking 
into account the uncertainties in the experimental data and material parameters. 
Varying the material parameters in the range of $\pm 10\%$ 
results in $\sim 20\%$ uncertainty in the anchoring energy. 
The influence of inaccuracies in the measured quantities (magnetic field, temperature,
thickness etc.) was essentially smaller ($\sim 3 \div 5\%$).

Thus, the temperature-induced orientational transitions planar $\to$ 
tilted $\to$ homeotropic at the
MBBA - $SiO$-evaporated substrate has been investigated.
From the experimental data on the transmitted light intensity in 
the presence of a magnetic field the 
temperature dependence of the surface angle has been found and the parameters
of the phenomenological model of surface energy have been determined.
The results demonstrate a good quantitative agreement
between the experimental data and the suggested surface energy model.

\vspace{12pt}
\noindent
{\bf Acknowledgments}

\noindent
We thank O. Khasanov for the help in setup construction and substrates preparation and
L. Kramer for fruitful discussions and critical reading of the
manuscript. 
A.K. is grateful to the Alexander von Humboldt Stiftung for equipment donation.
Financial support from INTAS Grant No. 96-498 is gratefully acknowledged.

\setlength{\baselineskip}{12pt}

\end{document}